\documentstyle[psfig]{mn}
\begin{document}

\title[Evidence of a decrease of kHz QPOs peak separation in 4U~1728--34]
{Evidence of a decrease of kHz QPO peak separation towards low frequencies in
4U~1728--34 (GX~354--0) }     
\author[Migliari, van der Klis \& Fender]
{S. Migliari\thanks{migliari@science.uva.nl}, M. van der Klis and 
R. P. Fender \\ 
\\
Astronomical Institute `Anton Pannekoek', University of Amsterdam,
and Center for High Energy Astrophysics, Kruislaan 403, \\
1098 SJ, Amsterdam, The Netherlands.\\
}

\maketitle

\begin{abstract}
We have produced the colour-colour diagram of all the observations of
4U~1728--34 available in the Rossi X-ray Timing Explorer public archive (from
1996 to 2002) and found observations filling in a previously reported `gap'
between the island and the banana X-ray states. We have made timing analysis
of these gap observations and found, in one observation, two simultaneous kHz
quasi-periodic oscillations (QPOs). The timing parameters of these kHz QPOs
fit in the overall trend of the source. The `lower' kHz QPO has a centroid
frequency of $\sim308$~Hz. This is the lowest `lower' kHz QPO frequency
ever observed in 4U~1728--34. The peak frequency separation between the
`upper' and the `lower' kHz QPO is $\Delta\nu=274\pm11$~Hz, significantly
smaller than the constant value of $\Delta\nu\sim350$~Hz found when the
`lower' kHz QPO frequency is between $\sim500$ and 800~Hz. This is the first
indication in this source for a significant decrease of kHz QPO peak
separation towards low frequencies. We briefly compare the result to
theoretical models for kHz QPO production.  

\end{abstract}


\section{Introduction}

Low-mass X-ray binary (LMXB) low-magnetic field neutron stars (NSs) 
often show pairs of kHz quasi-periodic oscillations (QPOs) in their power
spectra. The frequency separation $\Delta\nu$  
between these `twin peaks' is sometimes observed to be close to
the burst oscillation frequency $\nu_{burst}$, interpreted as
the spin frequency of the NS (Strohmayer et al. 1996; see Chakrabarty et
al. 2003), while sometimes it is close to half of it.    
Recent work on the millisecond pulsar SAX~J1808.4--3658 (where the spin
frequency is more straightforwardly measured, at 401~Hz; Wijnands et al. 2003)
indicates that in those latter cases $\Delta\nu$ is half the spin frequency. 
In some, perhaps all sources $\Delta\nu$ decreases significantly with
increasing QPO frequency. GX~17+2 shows a hint of a decrease at
lower frequencies (Homan et al. 2002), but is consistent with being constant.
In some sources, e.g. 4U~1728--34 (M\'endez \& van  
der Klis 1999), the peak separation is always significantly lower than the
burst oscillation $\nu_{burst}$, while in 4U~1636--53 (Jonker, M\'endez \& van
der Klis 2002) $\Delta\nu$ varies between lower and higher than
$\nu_{burst}/2$. 

Two competing interpretations exist for these high frequency timing features
(for a review see van der Klis 2000 and references therein) which make
definite predictions for $\Delta\nu$. The relativistic precession  
model (RPM: Stella \& Vietri 1998, 1999) is based on
free test particle orbits around a compact object. This model identifies the
`upper' kHz QPO as the frequency of an orbit in the disc and the `lower' kHz
QPO with the periastron precession of this orbit. The RPM model
predicts that the frequency separation $\Delta\nu$ decreases both when the kHz
QPO frequencies increase (as observed), but also when they decrease to
sufficiently low frequencies.    
In the sonic point beat frequency model (SPBFM: Miller, Lamb \& Psaltis 1998)
an orbiting clump at the sonic radius is related to the production of the
upper kHz QPO and the beat between this clump and the NS spin frequency
produces the lower kHz QPO. In this model $\Delta\nu$ should in principle be
equal to the spin frequency, but can decrease when the upper kHz QPO
increases due to the spiral-in of the clump.       
So, both theoretical models (RPM and SPBMF) can explain the observed decrease
of $\Delta\nu$ with increasing QPO frequency. What evidence can constrain or
test the validity of the models? The two models predict different behaviours
of $\Delta\nu$ as the kHz QPO frequencies decrease: the RPM predicts a
decrease of $\Delta\nu$, while the beat frequency model predicts a steady or
increasing (see e.g. discussion in Jonker et al. 2002) peak separation.
In addition, Osherovich \& Titarchuk (1999) developed the two-oscillator model
in which explained the kHz QPOs as arising from oscillations of hot blobs
thrown into the NS's magnetosphere. They identify the lower kHz QPO frequency
as the Keplerian frequency of the blob and the upper kHz QPO frequency as the
radial oscillator mode frequency, a non linear function of the Keplerian
frequency.  This model also gives a relation between the lower and the upper
kHz QPOs (dependent on the source; see e.g. Titarchuk, Osherovich \& Kuznetsov
1999) which can be tested by observations.   

In this Letter we present the timing analysis of recent observations of the
LMXB 4U~1728--34. We find twin peaks with lower kHz QPO frequency as low as
$\sim308$~Hz, and the first indication for a significant decrease of the kHz
QPO peak separation as the kHz QPO frequencies decrease.  If confirmed, this
evidence will strongly constrain theoretical models and, in particular, rule
out the sonic point beat frequency model in its current formulation.

\subsection{4U 1728--34}     
4U~1728--34 (GX~354--0; Forman et al. 1976) is a low-mass X-ray binary and
type-I X-ray burster (Lewin 1976; Hoffman et al. 1976), identifying the
accreting compact object as a neutron star. Hasinger \& van der Klis (1989)
classified 4U~1728--34 as an atoll-type X-ray binary.  Strohmayer et
al. (1996) found slightly drifting oscillations at $\sim 363$~Hz during X-ray
bursts, interpreted as the spin frequency of the NS. At high frequencies the
source shows kHz QPOs up to 1,200 Hz (Stromayer, Zhang \& Swank 1997) often in
pairs (lower and upper kHz QPOs). These pairs have a frequency separation
$\Delta\nu$ almost constant at $\sim350$~Hz (i.e. significantly lower than the
spin frequency) for lower kHz QPOs between $\sim500$ and $800$~Hz, decreasing
significantly to $\Delta\nu\sim280$~Hz above $800$~Hz (M\'endez \& van der
Klis 1999). No twin peaks with a lower kHz QPO frequency lower than $\sim
500$~Hz have been found in 4U~1728--34 until now.

\section{Observations and data analysis}

We have made timing analysis of 4U~1728--34 data from 1999 to 2002 performed
using the Proportional Counter Array (PCA) instrument on the Rossi X-ray
Timing Explorer (RXTE) satellite.    

\subsection{Colour-colour diagram}

\begin{figure}
\psfig{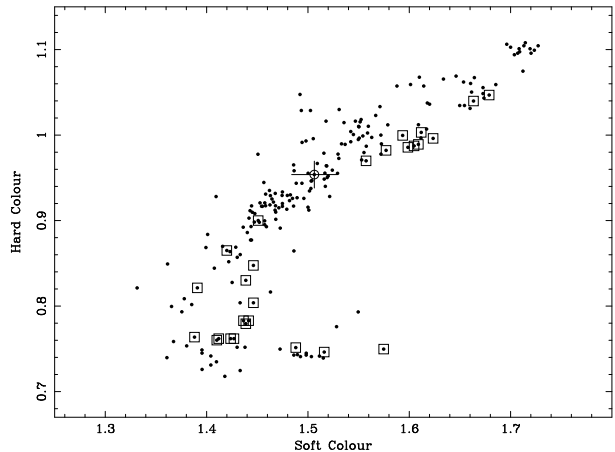}
\caption{X-ray colour-colour diagram (CD) of all the RXTE/PCA (PCU2 only)
observations of 4U~1728--34 available in the public archive: Soft
Colour=(3.5--6)~keV/(2--3.5)~keV, Hard
Colour=(9.7--16)~keV/(6--9.7)~keV. Plotted are the mean colours of each
observation. The squares mark the observation until 1997 analysed by Di Salvo
et al. (2000) and van Straaten et al. (2001); the circle marks the `gap'
observation which shows kHz twin peaks (with typical 90\% error bars
indicated)}
\label{CD}
\end{figure}

We have used {\tt Standard2} data of the Proportional Counter Unit 2 (PCU2,
always on in all epochs) to produce the CD of all the observations available
in the RXTE public archive (from 1996 to 2002).  The soft colour and the hard
colour are defined as the count rate ratio (3.5--6)~keV/(2--3.5)~keV and
(9.7--16)~keV/(6--9.7)~keV, respectively. We have normalised the colours of 4U
1728--34 to the colours of the Crab calculated with the closest observation
available to each 4U 1728--34 observation. In Fig.~\ref{CD} we show the mean
colours of all the observations of 4U 1728--34 (dots); squares are the
observations (1996--1997) previously analysed by Di Salvo et al. (2000) and
van Straaten et al. (2001); the circle marks the observation where we find
significant kHz twin peaks (see below).  Di Salvo et al. (2000) found a gap in
the CD between the island state and the banana state. This gap was only due to
lack of observations.  We concentrate on the observations filling in this
`gap' in between island and banana, namely in the range 0.9--0.98 of the hard
colour.

\subsection{Timing analysis}
For the production of the power spectra we have used {\tt event} data with a
time resolution of 125~$\mu$s. We rebinned the data in time to obtain a
Nyquist frequency of 4096~Hz. For each observation we created Leahy et
al. (1993) normalised power spectra from segments of 16~s length.
We removed drop outs and X-ray bursts from the data, but no background and
deadtime corrections were performed. We averaged the power spectra and
subtracted Poisson noise spectrum according to Zhang et al. (1995), shifted in
power to match the spectrum between 3000 and 4000 Hz, where there should be no
timing features. We then converted the power spectra to squared fractional
rms. We have fitted the power spectra with a multi-Lorentzian model (for
details see Belloni, Psaltis \& van der Klis 2001 and references therein). 
The power spectra of the gap observations are fitted using one broad
Lorentzian (L$_{b}$ or L$_{b2}$) to represent the low frequency noise and the
break frequency, one ( L$_{h}$) or two (L$_{b}$ and L$_{h}$) narrower
Lorentzians between $\sim10-50$~Hz, a broad Lorentzian around 100~Hz
(L$_{hHz}$) and one (except two observations in which two are needed,
i.e. with a single-trial significance $>3\sigma$; see \S~3) narrow Lorentzian
to fit the upper kHz QPO L$_{u}$ in the range $\sim500-700$~Hz (see van
Straaten, van der Klis \& M\'endez 2003 for details on terminology).

\begin{figure}
\psfig{figure=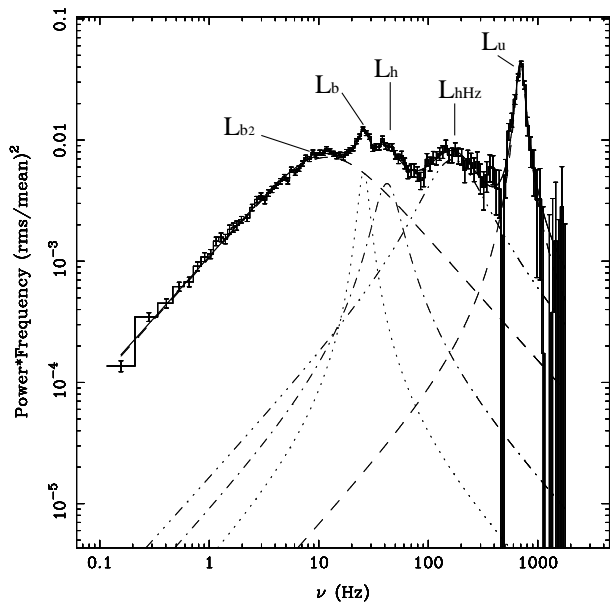,angle=0,width=8cm}
\caption{Typical power spectrum of a `gap' observation (i.e. observation
40033-06-03-01) with the best-fit multi-Lorentzian model.} 
\label{pow}
\end{figure}

\section{Results}
\begin{figure}
\psfig{figure=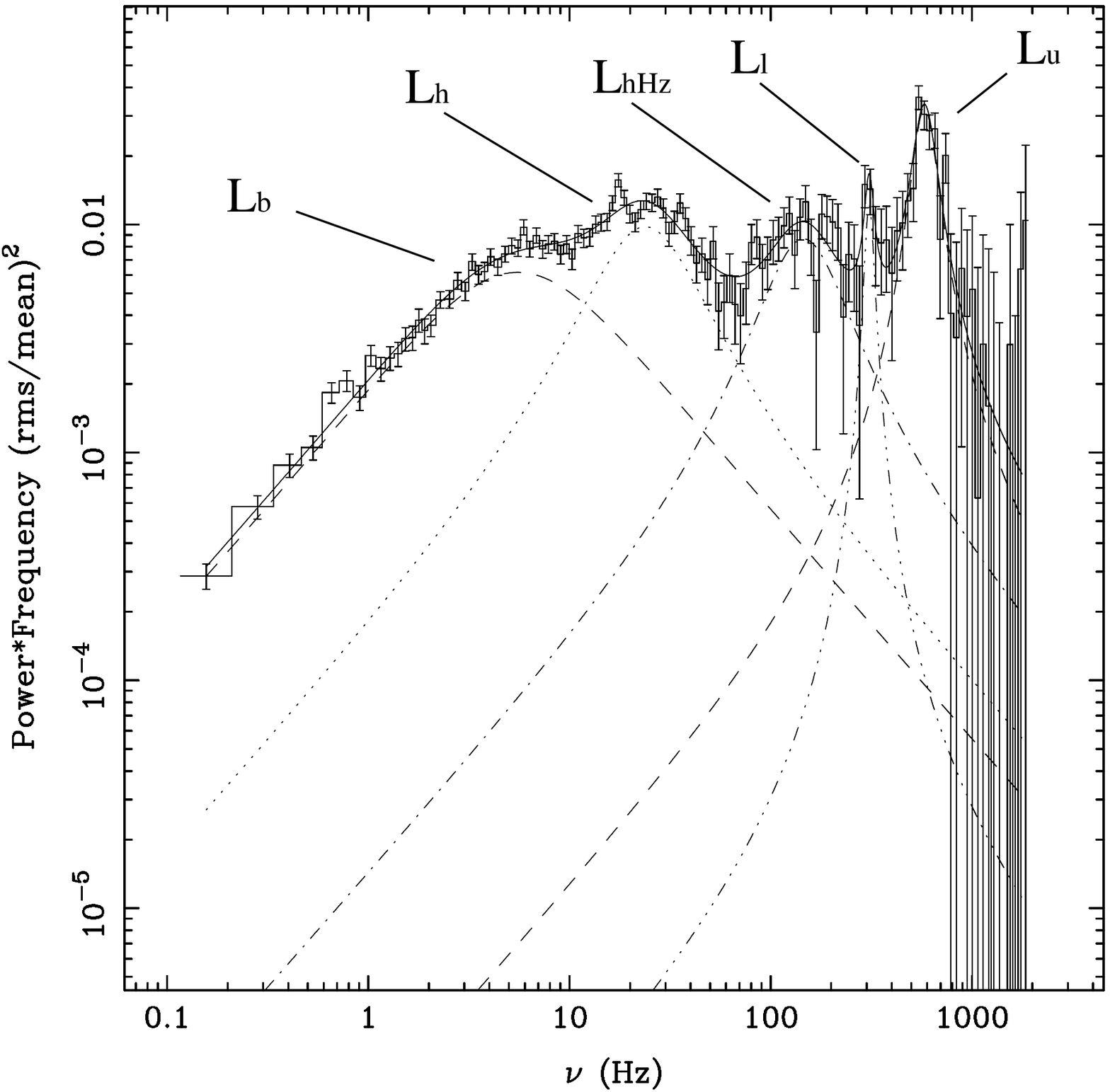,angle=0,width=8cm}
\psfig{figure=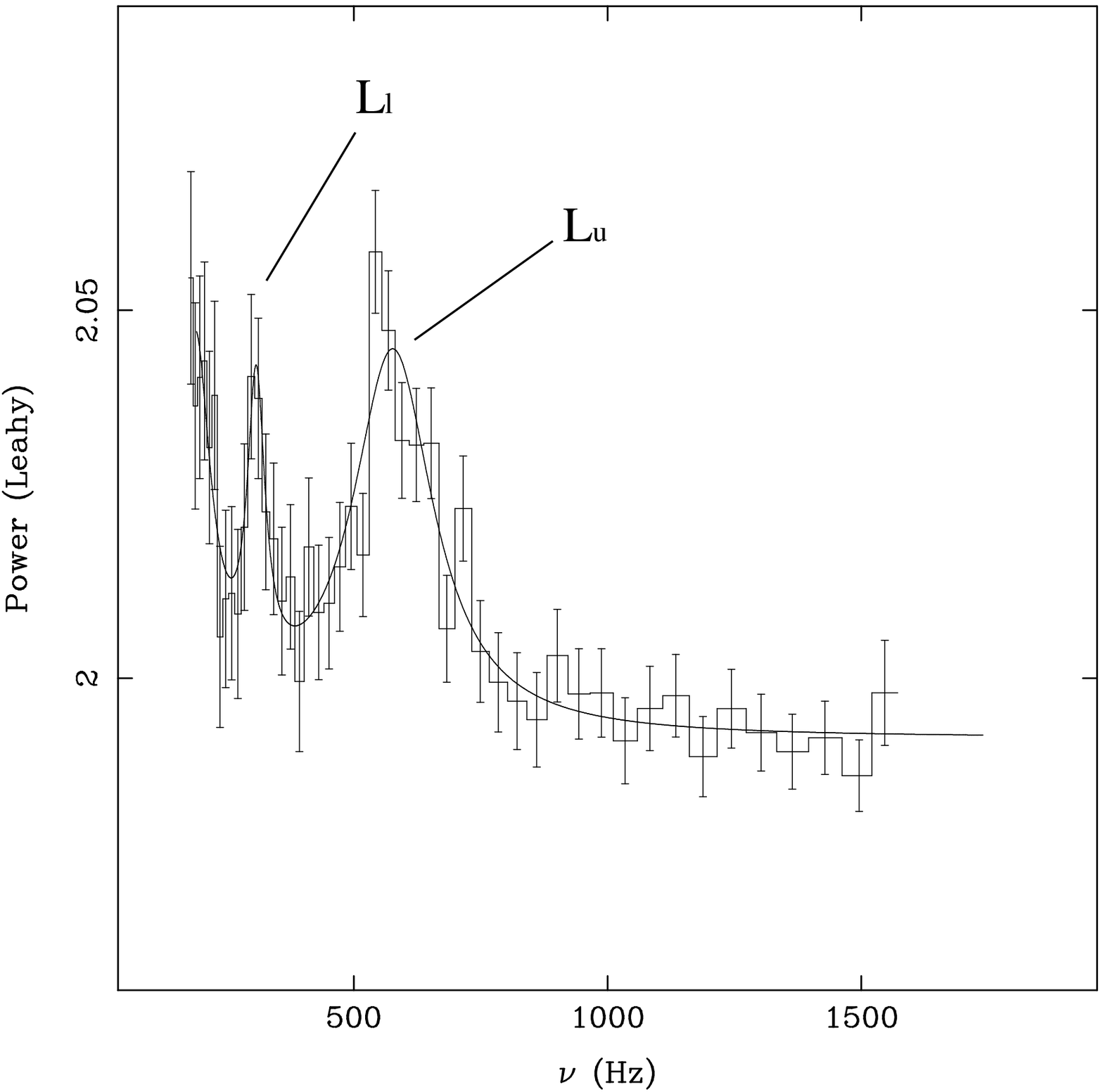,angle=0,width=8cm}
\caption{Top panel: power spectrum, in the power times frequency
representation, of the observation 50023-01-11-00 with the
best-fit multi-Lorentzian model. 
Bottom panel: high frequency range power spectrum of the same observation
showing the lower and upper kHz QPOs.}   
\label{twins}
\end{figure}

Fig.~2 shows a typical power spectrum of the observations analysed,
with the best-fit model. The upper kHz QPO is present in 46 out of 48 
observations. The power spectra are consistent with those
found around this position in the CD by Di Salvo et al. (2000) and van
Straaten et al. (2001; see Fig.~1, panel 7 in van Straaten et al. 2001). In
two observations, namely 50023-01-11-00 and 60029-02-02-00, we observe both
the lower (L$_{l}$) and the upper kHz QPO. 
In 50023-01-11-00  L$_{l}$ and L$_{u}$ are single-trial significant
(see \S~3.1) at a 4.4$\sigma$ and 7.9$\sigma$ level, respectively. In
60029-02-02-00 the significance is 
4.5$\sigma$ for the lower kHz QPO and 3.4$\sigma$ for the upper kHz QPO.      
Another observation (50023-01-31-00) shows a less significant
($2.3\sigma$) lower kHz QPO (the upper kHz QPO is significant at 10$\sigma$). 
In the observation 50023-01-11-00 (see Fig.~\ref{twins})
the centroid frequencies of the lower and the upper kHz QPOs are
$\nu_{l}=308\pm5$~Hz and $\nu_{u}=582\pm10$~Hz, respectively. The peak
separation is $\Delta\nu=274\pm11$. In the observation
60029-02-02-00 the centroid frequencies of the lower and the upper kHz QPOs
are $\nu_{l}=385\pm9$~Hz and $\nu_{u}=616\pm19$~Hz, respectively. The
peak separation is $\Delta\nu=231\pm21$.
To check for frequency variations in the kHz peaks we divided observation
50023-01-11-00 in to four time segments and observation 60029-02-02-00 in
to two segments, and analysed the power spectra of these segments. We note
that 60029-02-02-00 is `naturally' divided in two by an X-ray burst, hence in
this case we have analysed the pre-burst and post-burst power spectra. No
significant frequency variations are present in either observation. 

\subsection{Trials and significance of the kHz QPOs}

The number of gap observations analysed (observations with the hard colour in
the range 0.9--0.98 and with {\tt event} data with a time resolution of
125~$\mu$s) is 48. When we look for a timing feature,
the probability of a statistical fluctuation in the power spectrum at a
certain frequency increases with the number of trials, i.e. with the number of
power spectra searched, and with the range of
frequencies $\Delta\nu_{range}$ in which we expect the QPO. We estimate the
total number of trials as {\em number of power spectra} $\times
\Delta\nu_{range}/FWHM$. 
As reported in Fig.~3, in observation 50023-01-11-00 we have a very
significant upper kHz QPO at 582~Hz. We expect the lower kHz QPO (which has a
$FWHM\sim45$~Hz) in an interval of frequencies of about 200~Hz. This leads to
$\Delta\nu_{range}/FWHM\sim4.4$.  $4.4\sigma$ single-trial significance means
a chance probability of $1.08\times10^{-5}$ which multiplied by 48 trials and
by 4.4 gives $2.28\times10^{-3}$: the QPO is still significant at a $3\sigma$
level.  The kHz QPOs at $\sim385~Hz$ and $\sim615$~Hz of the observation
60029-02-02-00 turn out to be only marginally significant, below $3\sigma$. 
Therefore only one gap observation can be reported to show evidence of
twin peaks in the power spectrum.

\section{Discussion}

We have analysed the observations of 4U~1728--34, filling in the `gap' in the
CD (found by Di~Salvo et al. 2002) in between island and banana, namely in the
range 0.9--0.98 of the hard colour (see Fig.~1). In one of these observation
we find two significant kHz QPOs with a lower kHz QPO as low as 308~Hz (the
lowest `lower' kHz QPO ever found in this source). The lower kHz QPO has an
amplitude of $5.9\pm0.8$ per cent (rms) and Q=$\nu_{0}/FWHM=6.9\pm2.5$; the
upper kHz QPO has $12.8\pm0.8$ per cent rms and Q=$2.9\pm0.5$.  Do these
features fit in the overall timing behaviour of the source?  The amplitudes
vs. $\nu_{max}$ of the kHz QPOs of 4U~1728--34 are shown in Fig.~1 of M\'endez
\& van der Klis (2001). The lower and the upper kHz QPOs form two distinct
traces in the plot; the upper kHz QPO we find fits perfectly in the upper kHz
QPO trace. The lower kHz QPO appears discrepant, but there are no other lower
kHz QPO measurements below $\sim 600$ kHz with which to compare. 

\begin{figure}
\psfig{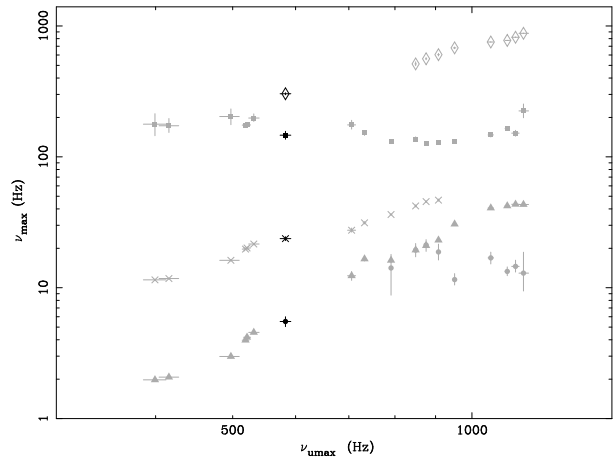}
\caption{Characteristic frequencies ($\nu_{max}$) of the Lorentzians used to
fit the power spectra of 4U~1728--34 and the $\nu_{max}$ of the upper kHz QPO
($\nu_{umax}$). The grey points are from van Straaten et al. (2002); the black
points are from the observation 50023-01-11-00. The circles
mark the L$_{b2}$, the triangles L$_{b}$, the crosses L$_{h}$, the
squares L$_{hHz}$ and the diamonds L$_{l}$.}  
\label{corr}
\end{figure}

Comparison of the upper kHz QPO frequency with Q and amplitude of both the
upper and lower kHz QPOs (see Fig.~4 in van Straaten et al. 2002) shows that
all the parameters of these QPOs fit in the overall trend of the source. In
Fig.~\ref{corr} we show the characteristic frequency $\nu_{max}$ (see Belloni
et al. 2001) of the upper kHz QPOs ($\nu_{umax}$) as a function of the
$\nu_{max}$ of all the other components used to fit the power spectra of
4U~1728--34. The grey points are from van Straaten et al. (2002) and the black
points are from the observation 50023-01-11-00.
The upper kHz QPO frequency of the observation 50023-01-11-00 is consistent
with the correlations found with all the other components and in particular
extend the L$_{l}$--L$_{u}$ correlation (stars) to lower frequencies. The
location of L$_{l}$ at $\nu_{lmax}=308\pm5$~Hz is too high for a hHz QPO which
in any case is separately detected (Fig.~3).  Whether or not the `gap' in the
L$_{l}$--L$_{u}$ correlation (diamonds in Fig.~\ref{corr}) between
$\sim600-800$~Hz is real must be further investigated. Clearly, bridging
this gap would increase our confidence in the identification of 
the peak at $\sim308$~Hz.

\begin{figure}
\psfig{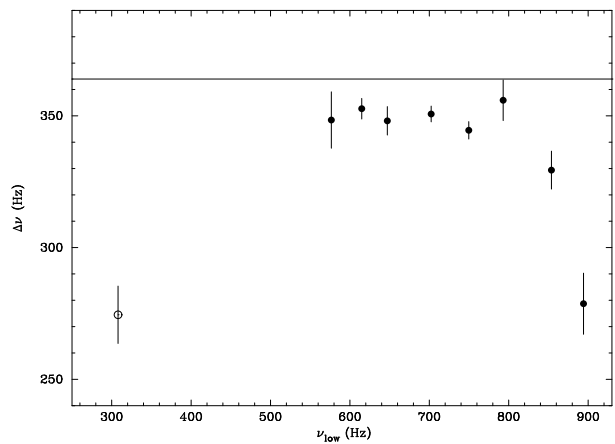}
\caption{Frequency separation between the lower and the upper kHz QPO
as a function of the frequency of the lower kHz QPO. The filled circles are
from M\'endez \& van der Klis (1999), the empty circle are the kHz peaks
we have found, the solid line indicates the
frequency of the burst oscillations at 363.9~Hz (Strohmayer et al. 1996).} 
\label{delta}
\end{figure}

The peak separation of the kHz QPOs we have found ($\sim 275$~Hz) is
significantly lower than both the spin frequency as inferred by burst
oscillations ($\sim 363$~Hz; Strohmayer et al. 1996) and the `saturation'
value (at $\sim349$~Hz) found by M\'endez \& van der Klis (1999) in the range
$\sim500-800$~Hz. In Fig.~\ref{delta} we show the peak frequency separation
$\Delta\nu$ as a function of the lower kHz QPO frequency. Black points are
from M\'endez \& van der Klis (1999) and the open circle is our
observation. The solid line indicates the frequency of the burst oscillations
at 363.9~Hz (Strohmayer et al. 1996). This indicates a decrease of the peak
frequency separation at low frequencies. This decrease is contrary to what is
expected from the SPBFM, which predicts a steady or increasing peak frequency
separation at low frequencies (see e.g. Fig.~2 in Lamb \& Miller 2001). A
clear prediction of a decrease of $\Delta\nu$ is made by the RPM. However, it
seems that the decrease at low frequencies we find is steeper than the
one predicted by the RPM: see e.g. Fig.~1 in Stella \& Vietri (1999). We have
also checked the relation between the upper and lower kHz QPOs as predicted by
the two-oscillator model. We have used the values of the constant parameters
of equation 15 in Osherovich \& Titarchuk (1999) as determined by Titarchuk,
Osherovich \& Kuznetsov (1999) for 4U~1728--34. For a low frequency QPO of
308~Hz the two-oscillator model predicts an upper kHz QPO of $\sim717$~Hz,
much higher than the one observed.



\section*{Acknowledgements}
SM would like to thank Steve van Straaten, Tiziana Di Salvo and Mariano
M\'endez for very helpful discussions. We would like to thank the anonymous
referee for her/his comments and suggestions.

\end{document}